\documentclass[twocolumn]{revtex4}
\usepackage{graphicx,epsfig}
\usepackage{amsmath}
\usepackage{amsfonts}
\usepackage[latin1]{inputenc}
\usepackage{natbib}
\usepackage{color}
\usepackage{url}
\usepackage{fancyhdr}

\usepackage{mathtools}

\newcommand{\bra}[1]{\left\langle #1 \right|}
\newcommand{\ket}[1]{\left|#1\right\rangle}

\DeclareMathOperator*{\maxZ}{max}

\newcommand{\gammabeta}{\{\boldsymbol{\gamma}, \boldsymbol{\beta}\}}

\begin{document}

\pagestyle{fancy}

\title{\Large Quantum Observables for continuous control of the Quantum Approximate Optimization Algorithm via Reinforcement Learning}

\author{Artur Garcia-Saez\footnote{Corresponding author: artur.garcia@bsc.es}}
\affiliation{Barcelona Supercomputing Center (BSC), Barcelona, Spain.}
\affiliation{Qilimanjaro Quantum Tech, Barcelona, Spain.}
\author{Jordi Riu}
\affiliation{Barcelona Supercomputing Center (BSC), Barcelona, Spain.}

% ----------------------------------------------------------------------------------------------------

\begin{abstract}
We present a classical control mechanism for Quantum devices using Reinforcement Learning. Our strategy is applied to the Quantum Approximate Optimization Algorithm (QAOA) in order to optimize an objective function that encodes a solution to a hard combinatorial problem. This method provides optimal control of the Quantum device following a reformulation of QAOA as an environment where an autonomous classical agent interacts and performs actions to achieve higher rewards. This formulation allows a hybrid classical-Quantum device to train itself from previous executions using a continuous formulation of deep Q-learning to control the continuous degrees of freedom of QAOA. Our approach makes a selective use of Quantum measurements to complete the observations of the Quantum state available to the agent. We run tests of this approach on MAXCUT instances of size up to $N=21$ obtaining optimal results. We show how this formulation can be used to transfer the knowledge from shorter training episodes to reach a regime of longer executions where QAOA delivers higher results.
\end{abstract}

% ----------------------------------------------------------------------------------------------------

\maketitle

% ----------------------------------------------------------------------------------------------------

\section{Introduction}

Optimization strategies performed by classical machines may help to overcome current limitations of NISQ devices \cite{nisq} --namely noise and decoherence time-- in order to solve hard computational tasks. Following this line of research, new algorithms such as the Variational Quantum Eigensolver \cite{vqe} and the Quantum Approximate Optimization Algorithm \cite{qaoa1} search for the optimal operations to be performed under experimental constraints. In this scenario, hybrid algorithmic proposals offer the possibility of using conventional optimization tools to optimally tune the parameter set controlling the Quantum device. This interface allows efficient state preparation of Quantum states which are solution to Quantum or classical problems formulated as a minimization task, delegating the optimization to an efficient classical algorithm.

The Quantum Approximate Optimization Algorithm (QAOA) \cite{qaoa1,qaoa2,qaoa_ions,qaoa_fermion,qaoa_perf,qaoa_continuous,qaoa_noise} provides a powerful circuit description \cite{qaoa_sup,qaoa_universal, qaoa_universal_2} that transforms a trivial Quantum state into one encoding the solution to a hard combinatorial problem. The circuit used by QAOA is constructed from the problem description and the solution is found as the ground state of a Quantum Hamiltonian adjusting gate operation by a set of real parameters. Finding these parameters through optimization provides a solution to the combinatorial problem, but at a large classical computational effort. A number of optimization techniques have been extensively used to attack QAOA classical optimization \cite{qaoa_qa,qaoa_perf,qaoa_gradient, qaoa_nn, qaoa_crooks,qaoa_transfer}.

We approach the optimization of hybrid algorithms as an interaction process between an active agent --with access to classical resources-- controlling the operation of a Quantum device executing a Quantum circuit, a scenario typically solved in Machine Learning with Reinforcement Learning techniques \cite{rl_book}. The agent learns from previous actions according to the rewards obtained and the effects that such actions had over the environment. After a training process, the agent exploits a learned strategy to maximize the reward received from the environment. The interaction is completely described by the reward function, the actions available to the agent, and the observations the agent may perform over the Quantum environment.

In this work we propose an approach to QAOA optimization based on continuous Reinforcement Learning. To complete this task we reformulate QAOA as an agent-environment interaction. A classical agent performs actions over a Quantum environment using a set of operations equivalent to the circuit formulation of QAOA. At each episode the agent chooses an action that modifies the Quantum state. The Quantum device performs the selected action, and returns a description of the Quantum state and a reward value. Measurement strategies are necessary to maximise the available information to the agent about the Quantum state. Together with the use of delayed rewards, this is the key contribution of our work. These observations are efficiently used by the agent to construct the optimal policy, and to find optimal reward values for large systems and circuits with numerically stability. Recent Reinforcement learning techniques based on Q-learning have shown great success in agent control over discrete actions \cite{rl_deep,rl_naf, rl_ddpg}. Our approach uses a formulation of Reinforcement Learning developed for a continuous action space \cite{rl_naf, rl_ddpg}. Using these techniques, we find optimal values of the objective function of the original QAOA problem after training, and we are capable to reach a regime of large circuit depth out of reach to global optimizers. Applications of Reinforcement Learning to Quantum systems have been explored extensively before in \cite{rl_fidelity,rl_quantum_control_1,rl_quantum_control_2}. Recently, approaches of Reinforcement Learning optimization of QAOA based on sampling \cite{rl_rigetti} or variations of the reward function \cite{rl_qaoa} have been applied to small Quantum systems.

We structure the present paper starting with the formulation of the QAOA in section \ref{section:QAOA}. Section \ref{section:RL} presents the tool set used for solving reinforcement Learning problems in a continuous action space. In section \ref{section:results} we plot results of our method that are summarized on section \ref{section:summary}.

\section{The Quantum Approximate Optimization Algorithm as a Quantum Environment} \label{section:QAOA}

\begin{figure}[h!]
\centering
\includegraphics[scale=0.25]{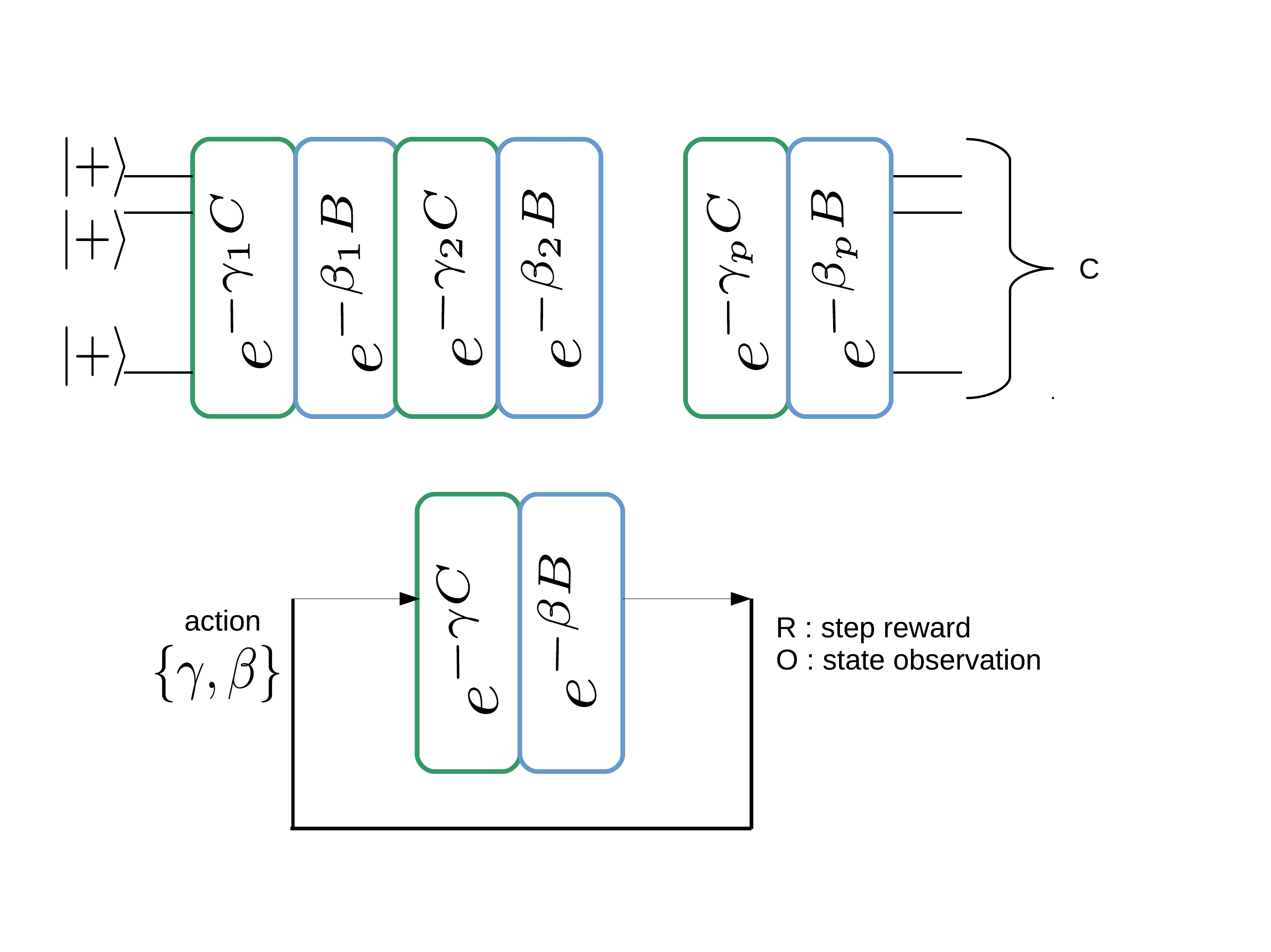}
\caption{Two equivalent formulations of the QAOA circuit. Upper panel: Block depiction of QAOA. At each step of the algorithm two continuous parameters $\{\gamma_i, \beta_i\}$ are used to perform an operation over a Quantum state initialized as $|s\rangle$. The final number of parameters $\gammabeta$ required to reach the final state $|\Psi\rangle$ is $2p$ for a given number of QAOA steps $p$. The objective function is evaluated after the last step. Bottom panel: Transformation of the QAOA algorithm into an interactive operational environment. At each step $i$, the initial internal state of the system $|\Psi_i\rangle$ is transformed by an action described with 2 continuous parameters $\{\gamma, \beta\}$, and returns a set of measurements $\{\langle O\rangle\}=\{X_i,Z_i,\ldots\}$ over the final state $|\Psi_{i+1}\rangle$ as an observation. The reward is obtained from Eq.\ref{eq:Hmaxcut} at the final step. We recover the original formulation of QAOA after $p$ steps and proper initialization on the first step to $|s\rangle$.}\label{fig:diagram}
\end{figure}

We study solutions to the MAXCUT problem, i.e. finding an optimal graph bi-partition maximizing the number of edges that connect the two partitions. The MAXCUT problem on a graph of size $N$ can be formulated in Hamiltonian form over a set of $N$ qubits as
\begin{equation}\label{eq:Hmaxcut}
C = \sum_{\langle i,j\rangle} \frac{1}{2}(1-\sigma^z_i \sigma^z_j) = \sum_{\langle i,j\rangle} C_{ij}
\end{equation}
with $m$ terms $C_{ij}$ accounting for adjacent edges in the original graph problem. The circuit implementation of QAOA \cite{qaoa1} requires a set of two operators $U(C,\gamma)$ and $U(B,\beta)$, constructed from the edge operator $C$ and the local operator $B = \sum\limits^n_{j=1} \sigma^x_j$:
\begin{equation}
\begin{aligned}
U (C,\gamma) &=& e^{-i\gamma C} = \prod\limits^{m}_{\alpha =1}\, e^{-i\gamma C_\alpha } \label{eq:eq3}\\
U (B, \beta) &=& e^{-i\beta B} = \prod\limits^n_{j=1}\, e^{-i \beta  \sigma^x_j}
\end{aligned}
\end{equation}
with $\gamma \in [0,\pi)$ and $\beta \in [0,2\pi)$. With $m$ terms in Eq.\ref{eq:Hmaxcut}, the algorithm requires a Quantum circuit depth $m p + p$. Operators $U(C,\gamma)$ and $U(B,\beta)$ are alternatively applied (see Fig.\ref{fig:diagram}) to a Quantum state initialized to the uniform superposition of computational basis states
$ \ket{s} = \frac{1}{\sqrt{2^n}}\ \sum\limits_z \ket{z} $. For an integer $p \geq 1$ and a set of $2p$ angles $\{ \gamma_1,  \beta_1 \ldots \gamma_p, \beta_p\} \equiv \gammabeta$, QAOA produces the Quantum state	
\begin{equation}
\ket{\boldsymbol{\gamma}, \boldsymbol{\beta}}_p = U (B, \beta_p) \, U (C, \gamma_p) \cdots U (B, \beta_1) \, U (C, \gamma_1)\, \ket{s}
\label{eq:eq7}
\end{equation}
which delivers a MAXCUT solution for large values of $p$ \cite{qaoa1,qaoa2}:
\begin{equation}
\lim_{p \to \infty} \left[ \maxZ_{\boldsymbol{\gamma}, \boldsymbol{\beta}} \bra{\boldsymbol{\gamma}, \boldsymbol{\beta}}_p C  \ket{\boldsymbol{\gamma}, \boldsymbol{\beta}}_p \right] = \maxZ C .
\label{FarhiEq11}
\end{equation}

We propose a revision of the QAOA optimization as an agent-environment interaction. A classical agent performs a sequence of actions over a Quantum device, and the combination of these actions is equivalent to the execution of QAOA. The actions transform the Quantum environment such that the final state is the result of the QAOA computation. Our approach is graphically expressed in Fig.\ref{fig:diagram}. A complete computation of QAOA is decomposed in a series of $p$ steps. A collection of $p$ steps, starting with the $\ket{s}$ state, forms an episode. At each step, an agent selects an action described by two real parameters $\{\gamma, \beta\}$, and receives a reward value and an observation of the current state of the Quantum system . This information is processed by the agent in order to construct a successful strategy that improves the final reward of the episode. The agent obtains information of the Quantum environment by a set of Quantum measurements $\{\langle O\rangle\}=\{X_i,Z_i,\ldots\}$. At the end of each episode the objective function Eq.\ref{eq:Hmaxcut} is evaluated and used as the final reward. However, different strategies to reward the agent along the episode can be designed, along the set of observations available at each step. This choice of observations and rewards will have deep impact in the strategies of the agent, and are a key ingredient in a successful Reinforcement Learning optimization.

\section{Reinforcement Learning in a continuous action space} \label{section:RL}

The general task of Reinforcement Learning {\cite{rl_book}} is to optimize the actions of an agent interacting with an environment in a Quantum state of which we have a collection of measurements $\{\langle O_i\rangle\}$. In our scenario, at each step the agent performs an action $\{\gamma,\beta\}$ that alters the state in the Quantum device $|\Psi_i\rangle$. The final goal is to maximize the total reward function $R_t = \sum_{i=t}^T d^{i-t}r(\{\langle O_i\rangle\},\{\gamma,\beta\})$ according to the rewards received $r(\{\langle O_i\rangle\},\{\gamma,\beta\})$ computed using Eq.\ref{eq:Hmaxcut}, and a discount factor $d$. At each time step $i$ in $[1,p]$ the agent chooses an action $\{\gamma,\beta\}$ according to its current policy 
$\pi(\{\gamma,\beta\} | \{\langle O_i\rangle\})$
and performs measurements to the current state and receives a reward.

We use an off-policy model-free Reinforcement Learning method presented in \cite{rl_naf} based on Q-learning. We define a policy-dependent $Q^ \pi$ function,
\begin{equation}
Q^\pi(\{\langle O_i\rangle\},\{\gamma,\beta\}) = \mathbb{E}(R_p |\{\langle O_i\rangle\},\{\gamma,\beta\}),
\end{equation}
such that the optimized policy is the greedy policy
\begin{equation}
\pi=\delta(\{\gamma,\beta\} = arg \max_{\{\langle O_{i+1}\rangle\}} (Q^\pi(\{\langle O_i\rangle\},\{\gamma,\beta\})).
\end{equation}

To obtain the $Q^\pi$ function from deep neural networks in a continuous action space, we use a decomposition of the network expression of $Q^\pi$ as the composition of an action term $A_{\theta_A} \equiv A(\{\langle O_i\rangle\},\{\gamma,\beta\}|\theta_A)$ and a value term $V_{\theta_V} \equiv V(\{\langle O_i\rangle\}|\theta_V)$
\begin{equation}
Q^\pi(\{\langle O_i\rangle\},\{\gamma,\beta\}|\theta_Q) = A_{\theta_A} + V_{\theta_V}
\end{equation}
Actions selected according to the greedy policy include a correlated random term modeled by an Ornstein-Uhlenbeck distribution in order to allow the agent a selective exploration of the action space.

\section{Results} \label{section:results}

We run simulations of a classical-Quantum system running the Q-learning algorithm, and report the reward at the end of each episode. We study the episode reward along a training session using an implementation of the Normalized Advantage Functions Q-learning algorithm \cite{rl_naf} with a noise contribution modeled by an Ornstein-Uhlenbeck distribution with $(\theta=0.01, \mu=0, \sigma=0.01)$. Values of $A$ and $V$ are computed and stored in neural networks with up to $20$ dense layers and a maximum of $256$ neurons at each layer. The agent receives a reward only at the final step of the episode, and the observation is restricted to a set of local operators $\{X_i, Z_i\}$. The learning rate is kept constant ($lr=10^{-4}$) along all executions. We have programmed the QAOA as a GymAI environment \cite{gym}, using Tensorflow \cite{keras-rl, tensorflow} to implement the NAF method, and QUPY \cite{qupy} to simulate the Quantum circuit.

\begin{figure}[ht]
\centering
\includegraphics[scale=0.6]{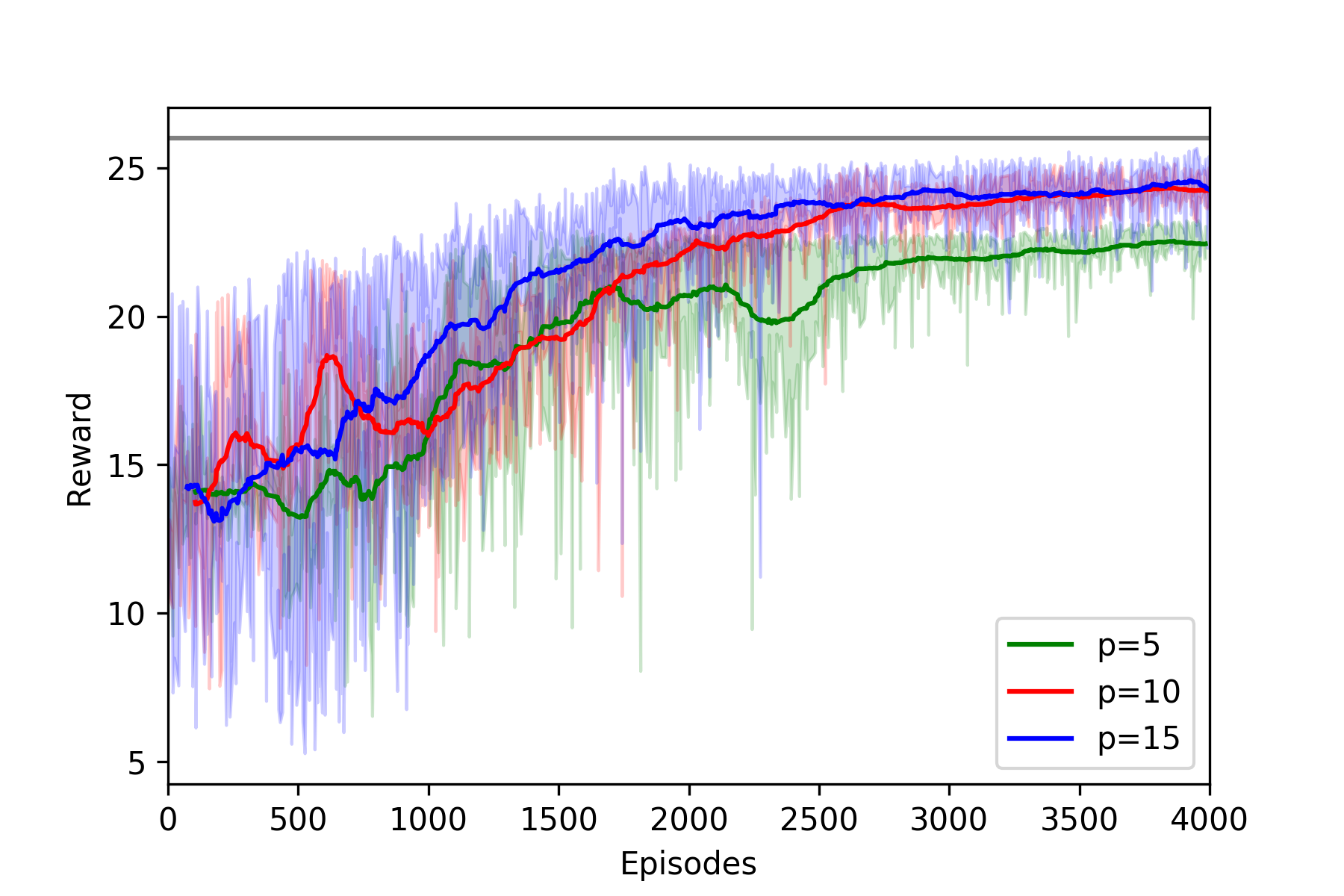}
\includegraphics[scale=0.6]{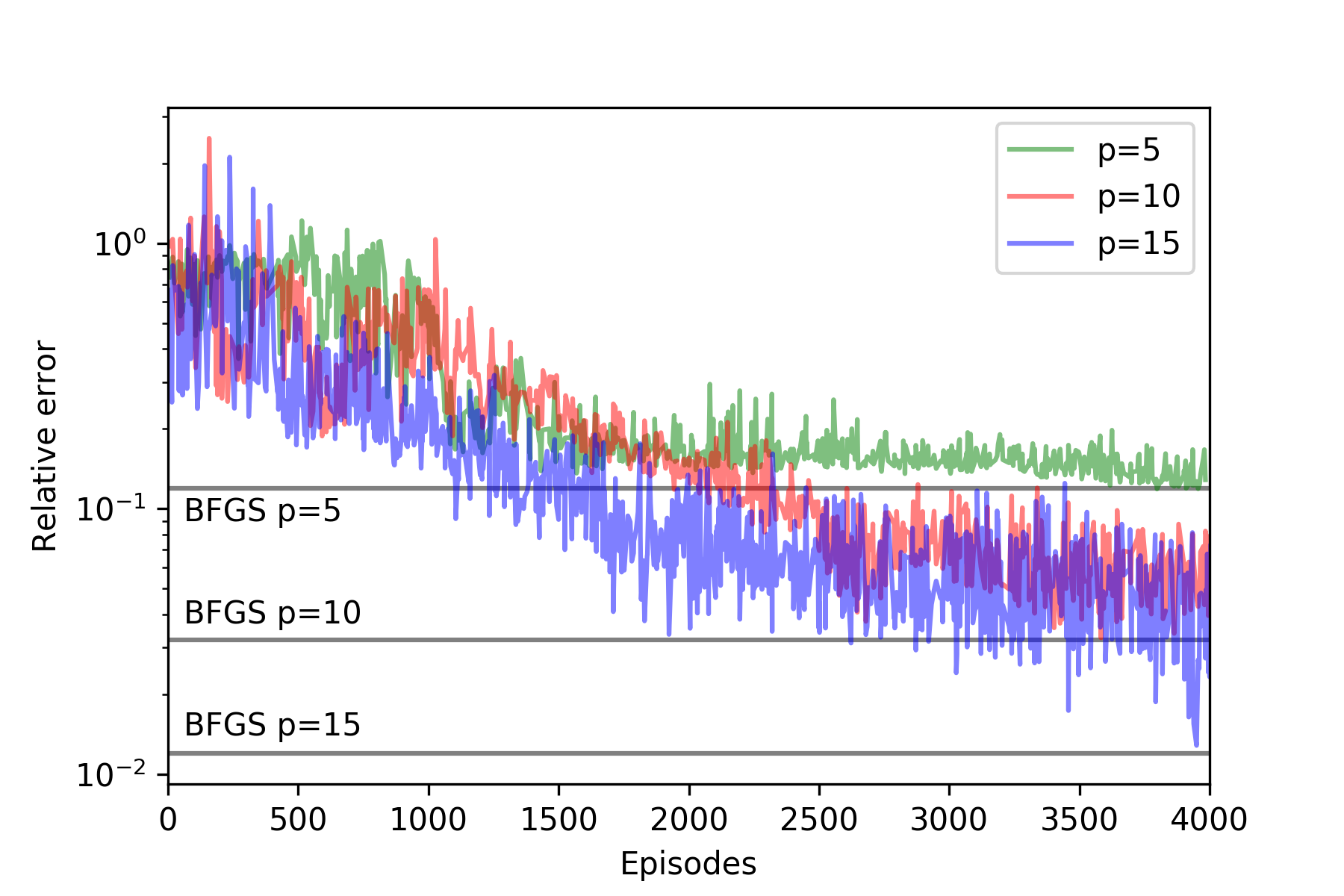}
\caption{Top: Reward computed by Eq.\ref{eq:Hmaxcut} at the end of each training episode for MAXCUT problems of size $N=13$. Mean rewards are averaged over several executions (included as a color line to guide the eye). An horizontal line marks the global optimal value. Episodes with a larger number of steps $p$ result in higher rewards, as expected from QAOA formulation. Bottom: Comparison of the relative error for each $p$ with the result of a global optimizer (BFGS) on the same instance. Consistent results are obtained with both methods after a similar number of evaluations.} \label{fig:fig2} 
\end{figure}

We pick an instance of MAXCUT defined by a random $3-$connected graph of size $N=13$ and show in Fig.\ref{fig:fig2} the reward at the end of each episode along the training. We observe that initially the reward obtained is consistent with an agent taking random actions, but evolves towards a set of parameters $\gammabeta$ delivering a solution the MAXCUT problem as the training proceeds. We show results for increasing values of $p$. The method exhibits a stable behaviour along a large number of training episodes, and according to QAOA condition 
\begin{equation}
\maxZ_{\boldsymbol{\gamma}, \boldsymbol{\beta}} \bra{\boldsymbol{\gamma}, \boldsymbol{\beta}}_{p'} C  \ket{\boldsymbol{\gamma}, \boldsymbol{\beta}}_{p'} \geq \maxZ_{\boldsymbol{\gamma}, \boldsymbol{\beta}} \bra{\boldsymbol{\gamma}, \boldsymbol{\beta}}_p C  \ket{\boldsymbol{\gamma}, \boldsymbol{\beta}}_p
\end{equation}
for $p'>p$ the method delivers better results for increasing values of $p$. In Fig.\ref{fig:fig2}(bottom) we compare results for $N=13$, $p= 5, 10, 15$ to a global optimization performed by a quasi-Newton method (BFGS) running during a comparable number of steps. 
Similar results are obtained for a collection of tests performed over $2-$ and $3-$connected graphs, for different values of $p$.

\subsection{Environment observations through local operators and reward strategies}

Reward functions play a critical role in the design of Reinforcement Learning strategies. At each step one can use the reward function to value the current step actions. A cumulative sum of partial rewards can be used as the final episode reward. However, our goal is the global optimization of the Hamiltonian in Eq.\ref{eq:Hmaxcut}. A reward value is returned only once the episode is finished, skipping limitations of local rewards that may lead to local minima of the global optimization. For the optimization of the QAOA we use delayed rewards motivated by the observation that global maximum of QAOA is not a combination of local solutions for smaller optimizations \cite{qaoa_perf}. In addition, a delayed reward strategy minimizes the evaluation of the reward functions at each step, reducing dramatically the number of executions for large $p$. 

We have reported here results where an agent has access to the Quantum state through a set of local measurements in different basis. At the end of each step these observations are used together with the reward received (if any) and the previous action taken by the agent throughout the learning phase. The set of observables define the observation space, and this can be complemented with additional information (e.g. the current number of steps). The observation available to the agent is limited to Quantum observables (i.e. access to measurements on a computational basis) along with limitations on the number of experimental repetitions of the experiment. Our particular choice of observations is inspired by the objective function Eq.\ref{eq:Hmaxcut} and the adiabatic principle: at the start of QAOA execution the initial state is fixed to $\ket{s} = \frac{1}{\sqrt{2^n}}\ \sum\limits_z \ket{z}$, an Eigenstate of the global $X$ operator. The final state is an Eigenstate of a diagonal Hamiltonian $C$ in the computational basis. To describe the state along QAOA evolution we choose the local observables $\langle X\rangle_i$ and $\langle Z\rangle_i$, $\forall i: 1\leq i\leq N$. The results shown in Fig.\ref{fig:fig2} include an evaluation of each of the terms $C_{ij}$, which has positive benefits for convergence. However, tests performed without these terms will deliver optimal results with a reduced number of measurements. This compromise between measurements and convergence has to be explored for each setting.

\subsection{Extending episode length to $p'>p$}

From the QAOA formulation we observe that reaching larger values of $p$ is a key element to obtain larger final rewards. Results reported in Fig.\ref{fig:fig2} are obtained after fixing $p$ at the start of the training process. This translates in slow converging rates for large $N$ and $p$, and may require fine adjustment of parameters along the training. Moreover, in this regime one may require the evaluation of larger neural networks where a cold start may complicate the optimization. We explore here a strategy based on incremental training to reach in a stable manner the regime of large $p$. An agent is initially trained in a $p$-length environment (i.e. all episodes have length $p$). After the training is completed to a converged reward, the agent is trained on longer episodes $p'>p$. The setup in the $p'$-training keeps the learned values from episodes of length $p$, i.e. the actions are computed from the same $A$ and $V$ functions. The agent improves the strategy with the $p'-p$ extra steps, potentially reaching higher rewards. This process is iterated to reach longer episode duration, a regime out of reach for global optimization techniques in our experiments.

\begin{figure}[h!]
\centering
\includegraphics[scale=0.6]{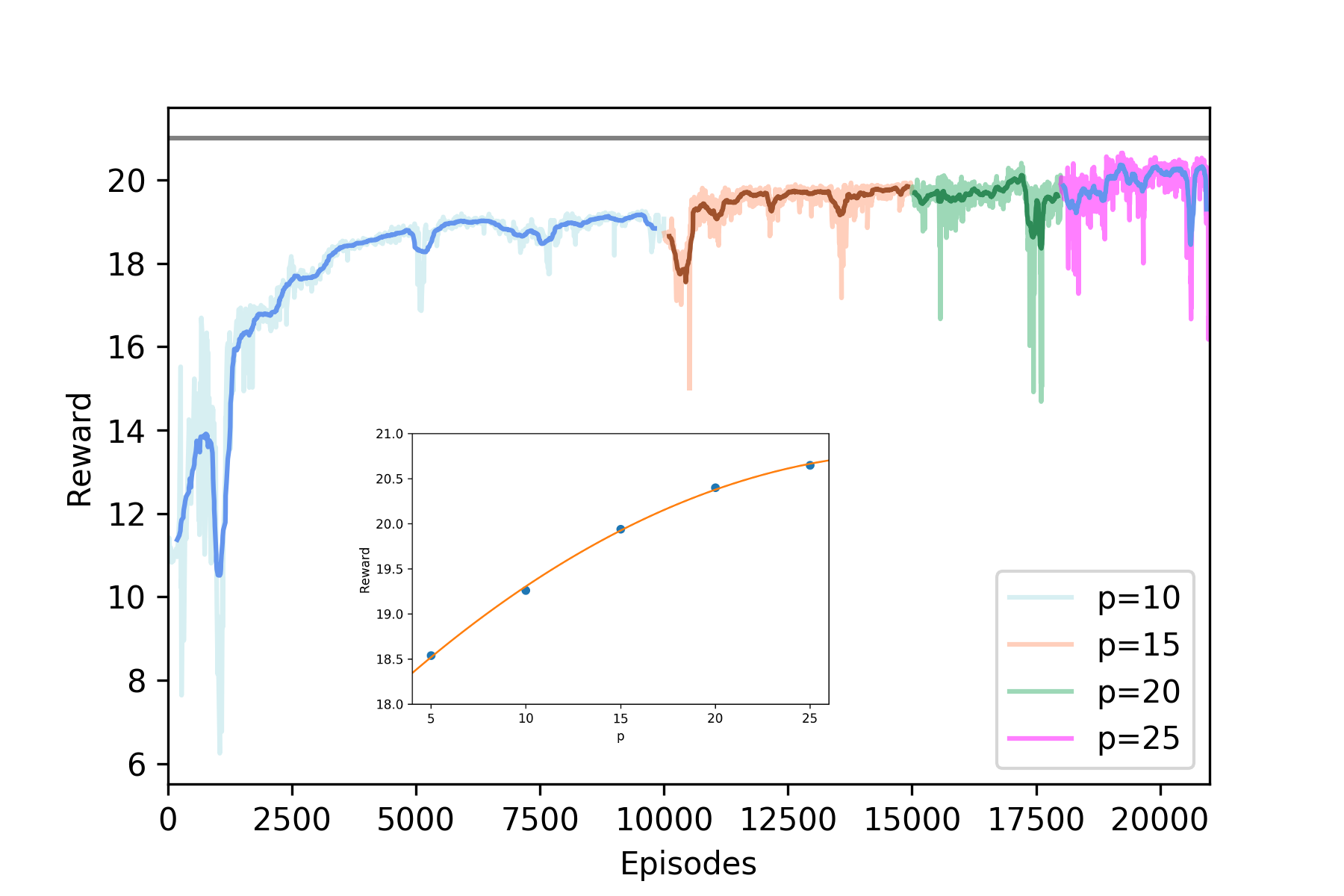}
\caption{Our strategy for using QAOA in a regime with a large number of steps $p$ reuse the training performed by the agent at a fixed episode length $p$ to start the training at episodes with $p'>p$. We plot here the results of successive training of a system of $N=21$ initially trained with $p=10$. After completion, data collected in this training phase is reused to solve the new episodes with $p=15,20$, progressively improving the results of the previous phase. We iterate again to reach training episodes with $p=25$. An horizontal line marks the global optimal value. Inset: evolution of the best reward obtained for each value of $p$.}\label{fig:fig3}
\end{figure}

We report in Fig.\ref{fig:fig3} results for a MAXCUT problem with $N=21$ and up to $p=25$. After an initial training with $p=10$, the agent is exposed to longer episodes with $p=15, 20$ and $p=25$. At the transition between training chapters where the value of $p$ is changed, instabilities appear initially as the agent optimizes the actions to the new setup. However, these instabilities are soon corrected and the agent reaches consistently larger rewards. Reaching larger values of $p$ is essential to a deep understanding of the computational capabilities of QAOA, as higher rewards are returned and eventually one may reach the exact solution. We show the maximum reward returned for values of $p=5,10,15,20,15$ in the inset of Fig.\ref{fig:fig3}.

\section{Summary} \label{section:summary}

We have formulated the QAOA as an agent-environment interaction where a classical agent controls a Quantum device. Applying Reinforcement Learning strategies to QAOA we show how to interactively drive the execution of Quantum algorithms in order to solve hard combinatorial problems. Our agent controls the optimization having access to a selection of information about the performance of the algorithm available through Quantum measurements, a collection of data that extends the capabilities of global optimizators relying solely on the value of the objective function. Our results suggest that this information is a valuable resource in running-time optimization, and that Reinforcement Learning methods used in our work are able to exploit the advantage provided by these observations.

We have explored the QAOA parameter space by means of Reinforcement Learning methods formulated for a continuous action space. 
Our work allows a general revision of Quantum algorithms using the corpus of Reinforcement Learning techniques developed in the last years, and successfully applied to a wide range of problems in robotics and optimal control. Numerical stability allows us to reach a regime of large $p$ where QAOA delivers increasingly higher rewards. The tools developed here contribute to understand how QAOA performs in a large range of $p$ values.

A fundamental difference in the execution of our method would appear running the Reinforcement Learning algorithm in a real Quantum device. For a given intermediate step $k$ of an episode, the Quantum state is repeatedly prepared to obtain the set of observed quantities. This requires a memory expression of the previous steps in terms of classical parameters, i.e. $\{\gamma,\beta\}_i, \forall i < k$. This little overhead results in a longer state preparation for later steps on each episode. However, this overhead can be reduced by grouping more actions inside each episode, or equivalently reducing the number of Quantum observations.

% ----------------------------------------------------------------------------------------------------

\section*{Acknowledgement}

We acknowledge funding from project FIS2017-89860-P (MINECO/AEI/FEDER, UE) and support of NVIDIA Corporation for this research.

% ----------------------------------------------------------------------------------------------------

\end{document}